\documentclass[journal]{IEEEtran}
\usepackage{color}
\usepackage{cite}
\usepackage{textcomp}
\usepackage{xcolor}
\usepackage{balance}
\usepackage{graphicx}
\usepackage[english]{babel}
\usepackage{amsthm}
\usepackage{amsmath,amssymb}
\usepackage{mathtools}
\usepackage{tikz}
\usepackage{pgf}
\usepackage{amsfonts}
\usepackage{bm}
\usepackage[bottom]{footmisc}
\usepackage[ruled,linesnumbered]{algorithm2e}

\usepackage{setspace}
\usepackage{subcaption}
\usepackage{array}
\newtheorem{definition}{Definition}[section]

\DeclareMathOperator*{\argmax}{arg\,max}
\SetKwInOut{KwIn}{Input}
\SetKwInOut{KwOut}{Output}
\newtheorem{rem}{Remark}

\newtheorem{proposition}{Proposition}
\allowdisplaybreaks

\def\BibTeX{{\rm B\kern-.05em{\sc i\kern-.025em b}\kern-.08em
    T\kern-.1667em\lower.7ex\hbox{E}\kern-.125emX}}
\begin{document}
\title{Double reflections Assisted RIS Deployment and Energy-efficient Group Selection in mmWaves D2D Communication}


\author{Authors}
\author{Lakshmikanta Sau, \textit{Member, IEEE,} and Sasthi~C.~Ghosh
\thanks{L. Sau and S. C. Ghosh are with the Advanced Computing \& Microelectronics Unit,  Indian Statistical Institute, Kolkata 700108, India. (E-mail: lakshmikanta\_r@isical.ac.in, sasthi@isical.ac.in).
}}

\maketitle
\begin{abstract}
     Reconfigurable intelligent surfaces (RISs) offer a viable way to improve the performance of the multi-hop device-to-device (D2D) communication. However, due to the substantial propagation and penetration losses of the millimeter waves (mmWaves), a direct line of sight (LoS) link and close proximity of a device pair are required for a high data rate. Static obstacles like trees and buildings can easily impede the direct LoS connectivity between a device pair. Hence, RIS placement plays a crucial role in establishing an indirect LoS link between them.
 Therefore, in this work, we propose a set cover-based RIS deployment strategy for both single and double RIS-assisted D2D communication. In particular, we have demonstrated that permitting reflections via two consecutive RISs can greatly lower the RIS density in the environment, preventing resource waste and enabling the service of more obstructed device pairs. After the RIS deployment, for information transfer, we also propose an energy-efficient group selection criteria. Moreover, we prove that sometimes double reflections are more beneficial than single reflection, which is counter-intuitive.
  Numerical results show that our approach outperforms a random and a recent deployment strategy.
\end{abstract}
\section{introduction}
Due to the exponential increase in the number of users and the demand for a high data rate, there is a heavy crisis in bandwidth \cite{ericsson}. Device-to-device (D2D) \cite{D2D_communication} communication is one of the prominent solutions for this need. In D2D communication, two proximity users can directly communicate with each other without the help of a base station. Note that we can deliver a high data rate over a limited distance in millimeter waves (mmWaves) D2D communication \cite{mm1}. However, in mmWaves D2D communication, penetration loss is very high due to the presence of randomly located obstacles and trees \cite{panetration_loss}. In this context, over the past few decades, technologies called beamforming have been developed to address this need for increased data rates \cite{beamforming}. Regardless of the technical details, the common goal of all of them is to intelligently adapt to the randomly fluctuating wireless channel rather than to control it. Therefore, the so-called reconfigurable intelligent surface (RIS)\cite{risi1} is a novel technology that claims to solve this problem. An RIS consists of an array of reconfigurable passive elements embedded in a flat metasurface \cite{ris2}. Every passive element contains a series of embedded PIN diodes that may be switched between the ON and OFF states by adjusting the biased voltage using a direct current (DC) input line. RISs can change a propagation environment into a desired form \cite{risi2}. As RIS reflects an incident signal in a desired direction, it does not need any radio frequency chains. The main aim of using RIS is to give an indirect line-of-sight (LoS) path through RIS to blocked device pairs and it also reduces hardware cost \cite{opris}. It is noted that an RIS can be used as an amplify and forward (AF) relay, that is, it reflects all signals with an amplification factor $1$. Since RISs consist of many patches, grouping strategies have been proposed by several works \cite{grouping,partition2,partition3} to minimize the significant channel estimation overhead in RIS-assisted systems. The authors in \cite{traing_and_reflection_pattern} investigate how to minimize the mean-squared error of the channel estimation by jointly optimizing the training signals at the user equipments and the reflection pattern at the RIS.

Nowadays, RIS-assisted D2D communication forms a new research direction \cite{traing_and_reflection_pattern, ran, Distributed_RIS}, and the majority of recent research focuses on RIS usage by assuming that RISs are already set up and accessible. 
A novel RIS deployment strategy in the presence of phase errors is developed in \cite{Distributed_RIS} by keeping in mind the fair and acceptable restriction of the total number of RIS elements.  However, most of the recent research on RIS deployment techniques assumes single reflection only \cite{opris, ran, kishk2020exploiting,sauris,deb2021ris}. Note that a few device pairs may not be able to communicate with each other via a single reflection, depending on the shape and position of the obstacles. In practice, the secondary reflections are not insignificant, especially in metropolitan settings where the RISs are not widely dispersed. Therefore, in a few cases, double reflections can be impactful when a single reflection is insufficient for RIS-assisted communication. Additionally, by allowing double reflections, we can serve more obstructed device pairs without deploying more RIS.  A few studies have taken double reflection into consideration \cite{mhop1,2hop,risi3,dramp}, but they assumed that the RISs were already deployed rather than developing any deployment strategy for double reflection.

In this work, we ignored the role of triple and higher order reflections due to significant effective path loss \cite{2hop,dfaf}. Unlike passive RISs that reflect signals without amplification, active RISs can amplify the reflected signals via amplifiers integrated into their elements. However, it requires high power consumption, and it needs integrated power supplies, amplifiers, and control circuits \cite{active_vs_passive_ris}. In this work, we use the passive RISs for power-constrained scenarios \cite{dramp}. Here, we first develop a set cover based RIS deployment strategy that helps to get an indirect LoS link between a transmitter and receiver pair in order to establish communication when there is no direct LoS link between them by considering both single and double reflection. After deploying the RISs, if a device pair wants to communicate with each other, multiple RIS may be available to assist their communication. To reduce the effective number of RIS elements and the training overhead, the authors of \cite{traing_and_reflection_pattern} have investigated a subgroup-wise joint training and optimization of the reflection pattern that divides all RIS elements into multiple subgroups, each of which is made up of a collection of nearby RIS elements that share a common reflection coefficient. Note that, due to the cost of channel training and feedback overhead, the entire RIS consumes significantly more energy as compared to one of the subgroups. Accordingly, in a specific scenario, multiple subgroups may exist that can provide the required data rate to the requesting pair. In such a case, without using the entire RIS, we select a specific energy-efficient subgroup for information transfer among all available RISs. More specifically, our contributions are as follows. 

\begin{itemize}
    \item To the best of our knowledge, this is the first work that considers both single and double reflections for deploying the RISs. Due to the obstacles's shape and density, few device pairs may not get indirect LoS via single reflection. In such cases, double reflection can help them to get an indirect LoS. This in turn increases the number of served blind pairs. 
    Moreover, double reflection has also been demonstrated to lower the number of RIS needed.
  \item Additionally, we have mathematically proved that under certain conditions, double reflections are more energy-efficient than single reflection, which is quite counter-intuitive. 
    \item We have demonstrated that three reflections may provide only a slightly higher throughput, but the corresponding power consumption is reasonably high. As a result, triple or higher reflections may not be as good as double reflection in terms of energy efficiency. This further justifies the widespread assumptions of ignoring the triple and higher order reflections in the literature.
    \item It is shown that the approximation ratio of our set cover based RIS deployment algorithm is  $O(\log |B|)$, where $|B|$ represents the total number of blind pairs. This ratio is the best possible, as no polynomial-time algorithm for set cover can give a better approximation ratio than $O(\log n)$ unless P = NP.
    \item Each RIS is divided into a number of non-overlapping subgroups to reduce the channel estimation overhead. Subsequently, we propose an energy-efficient algorithm to find the appropriate subgroups of an RIS to achieve higher sum throughput.
\end{itemize}
Simulation results demonstrate that our proposed strategy can increase sum throughput significantly and reduce the number of overall RIS deployments in comparison to a random \cite{ran} as well as a recent deployment strategy \cite{opris}.

This paper is organized as follows. In Section \ref{rwork}, we have introduced the related and existing works. We have discussed the system model and preliminaries in Section \ref{sys}. Mathematically, the problem formulation is discussed in Section \ref{prob}.  The strategic deployment of RISs, the group selection strategy, and their analysis are discussed in Section \ref{prop}. Thereafter, in section \ref{simu}, we have discussed the simulation results and compared this with the existing placement strategies. Finally, in Section \ref{con}, we give concluding remarks and the future direction of the research.
\vspace{-2mm}
\section{Related Work}\label{rwork}
Numerous recent studies have examined the effects and advantages of RISs in mmWave D2D communication. The authors in \cite{risi3} focus on the uplink of an RIS-assisted D2D-enabled cellular network. Additionally, in \cite{mm1}, the authors discussed the utilization of high-frequency signals like mmWaves to achieve the goal of producing high-speed data rates for short-distance communication. Due to the randomly located obstacles, there is a heavy path loss in mmWave wireless communication. As a result, if a direct LoS link between a device pair does not exist,  we need to bypass the signal. The signals can be bypassed using a single reflection via one RIS or reflections via more than one consecutive RISs.  Therefore, RIS-assisted communication is one of the ways to achieve an indirect LoS link between an obstructed device pair \cite{mmris1,dar}.

 The authors in \cite{opris} have provided simulation evidence on how locating RISs near users enhances performance by optimizing the RIS orientation and horizontal distance in a single base station and single-user downlink network. The authors in \cite{tgcn_6g}, have discussed the most recent RIS technology enabling methods and the significant advantages that come with them from the perspectives of RIS modulation and RIS reflection in 6G networks. The perspectives on how RISs can be incorporated into commercial networks and the corresponding standards are discussed in \cite{RIS_industry}. In \cite{kishk2020exploiting}, the authors investigate how the performance of cellular networks is affected by the widespread deployment of RISs using tools from stochastic geometry by considering the exploitation of the randomly located obstacles. However, after random deployment, there are still many device pairs that can not communicate with each other. Motivated by this, a graph theory based strategic placement of RISs is developed in \cite{sauris} by assuming a single-RIS-assisted communication framework. Here, the total number of RISs used and the deployment cost are reduced by providing indirect LoS via the strategically placed RISs. A placement strategy using the set cover approach for single RIS-assisted communication has been described in \cite{deb2021ris}. Note that all these placement strategies are applicable to single RIS-assisted communication only. However, a few device pairs can not communicate with each other via a single reflection depending on the shape and position of the obstacles. In this context, if double reflections are allowed, some of them may be able to communicate with each other. There are few works that consider double reflections as well. The work in \cite{2hop} shows that the multi-RIS double reflections may be used to considerably increase the communication range by properly tuning the RISs. In order to increase the data rate, the authors in \cite{risi3} used the ability of the RISs to change the phase shifts of the elements and provide advantageous beam steering considering double reflection. To increase the coverage area, a multi-hop RIS-assisted communication framework is developed in \cite{dramp} by considering both single and double reflection. However, all these secondary reflection based works assume that the RISs are already deployed.

It is clear that RISs are put to innovative use in a variety of ways to improve the caliber of wireless communication services. More specifically, in a D2D communication context, RISs are used to reduce blind pairs, eliminate interferences, and avoid obstacles.
However, most of the recent studies assume that the RISs are already deployed either randomly \cite{ran} or strategically in single-reflected RIS communications \cite{opris,wide_deployment}. Also in the multi-hop scenario, where secondary reflections are allowed, they also assumed that RISs are already deployed. Furthermore, for a large RIS, we have to compute a huge number of channel estimations. As a result, the authors in \cite{partition3,pm_dcsk} and \cite{Tnse} discussed a novel RIS grouping strategy to reduce the channel estimation overhead. The study in \cite{2hop} and \cite{ec4} examines how RISs can improve energy-efficiency in mmWaves D2D communication. Moreover, wireless communication is greatly impacted by the strategic placement of RISs where double reflections are allowed. Our goal in this work is to strategically position a minimum number of RISs considering both single and double reflections, and select an energy-efficient subgroup to improve network performance.

\vspace{-2mm}
\section{System Model and Preliminaries}\label{sys}

\subsection{Network Topology}
Consider a wireless communication system that operates in a rectangular area that is partitioned into small squares or grids of unit size, which is shown in Fig. 1. The rectangular area consists of $M$ rows and $N$ columns. We assume that each unit square grid is identical with respect to their center and we also assume that the center of the leftmost corner square is the origin. It is noted that the device that will operate in this setting will have highly directed antennas. The position of a user is approximated to the center of the grid within which it lies. We assume that a device can communicate with another directly if there is a LoS link between them and they are within a threshold distance $r$. This is because, as the signal gain diminishes with increasing distance, it will become insufficient for communication beyond a certain range.

Now, we are presenting a few definitions below, which will be used throughout our discussion.

\vspace{-2mm}
\begin{definition}[Direct LoS link]
 If there is no blockage between two users $u$ and $v$  and they are within a distance $r$ then the link between them is called a direct LoS link  \cite{kishk2020exploiting}.
\end{definition}
\vspace{-2mm}
\begin{definition}[Blind pair]
    A device pair $(u,v)$ lies within a distance $r$ is said to be a blind pair if there is no direct LoS link between them.
\end{definition}
\vspace{-2mm}
\begin{definition}[single reflection]
    A device pair $(u,v)$ is said to be coverable via single reflection if there is no direct LoS link between $u$ and $v$ but they are connected via an intermediate RIS, i.e., there is a direct LoS link between $u$ to an RIS and that RIS to $v$.
\end{definition}
\vspace{-2mm}
\begin{definition}[Double reflections]
The device pair $(u,v)$ lie within a distance $r$ is said to be coverable via double reflections if $u$ and $v$ can communicate with each other in two-hop using two consecutive RISs $R_i$ and $R_j$, i.e., there is a direct LoS link between $u$ to $R_i$, $R_i$ to $R_j$, and $R_j$ to $v$, respectively.   
\end{definition}
\vspace{-2mm}
\begin{definition}[Coverable blind pairs]
     If a device pair $(u,v)$ is coverable via single reflection or double reflections, or both of them, then it is considered as a coverable blind pair, else it is called a totally blind pair. 
\end{definition}
\vspace{-3mm}
\subsection{User and Obstacle Characteristics}\label{umob}
We consider all the devices in this communication scenario to be pseudo-stationary \cite{singh2019mobility}, i.e., during communication time, a device does not move outside the grid. We also assume that a device follows any mobility model within a grid at any time instance. However, we presume that a device's location in a unit grid is roughly determined by the grid's center. It is noted that if we consider a device is within a grid that means it lies at the center of the grid.
\begin{figure}
    \centering
    \includegraphics[width=0.45\linewidth]{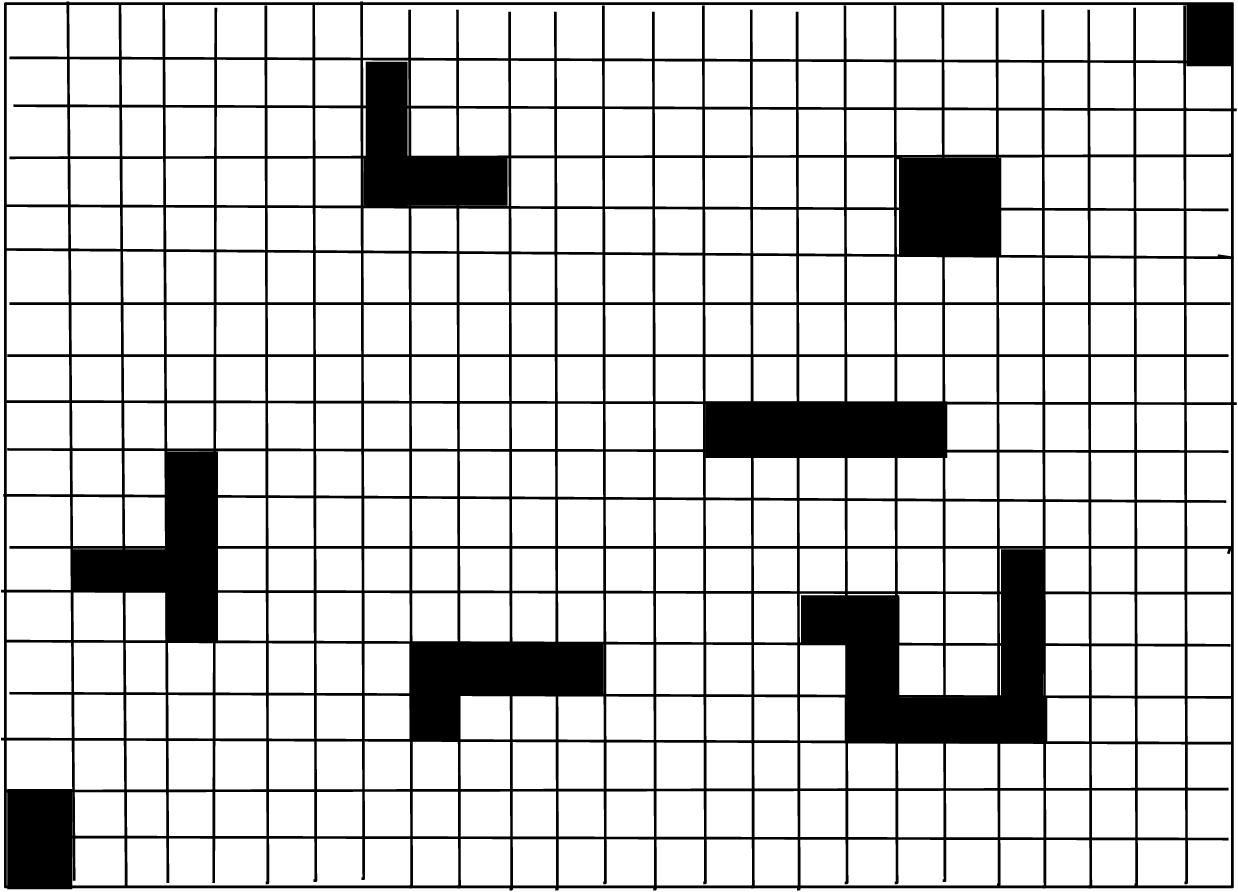}
    \caption{Grid and obstacle model}
    \label{gmod}
    \vspace{-6mm}
\end{figure}

 Here, we position the obstacle inside the grids following the acquisition of the satellite images.
 We assume that the satellite images give the proper position of the obstacles. In Fig. \ref{gmod}, we consider that the black cells represent the location of the obstacles. We assume that, if a grid/cell is blocked, it means the whole cell is blocked and there will not be any partially blocked cells. Multiples of these blocked zones combine to build a polygon that closely resembles the shape and size of the obstruction. Within a block cell, a device can not lie, i.e., a device can lie only within the free cells. It is noted that an RIS can not be placed within a block cell, it will be strategically placed on poles in the free zones. For simplicity and mathematical tractability, we discretize the service area into unit-size grids as in \cite{deb2021ris} and thus approximate the real-life 3D environment to a 2D plane. 
\vspace{-2mm}
 \subsection{RIS Grouping}
 Obstacles prevent a device pair from directly communicating with one another. Therefore, to provide an indirect LoS between them, we use single and double reflections via one and two consecutive RISs. Let us assume that an RIS $R_i$ consists of $M_r$ rows and $N_r$ columns, which are effectively controlled to adjust both the amplitude and phase of the incident waveform. However, only the phase is tuned or optimized, and the amplitude factor is fixed to unity for the purposes of simplicity and mathematical tractability \cite{partition3}. We use a grouping approach in order to minimize the channel estimation overhead \cite{partition2}. Furthermore, for a large RIS, the phase shift computation is hard, and a minimum energy is required for each phase shift. Therefore, we divide $R_i$ into $K_g$ number of non-overlapping subgroups $R^t_i$ to minimize the energy consumption where $1\leq t \leq K_g$, and each $R^t_i$ consists of $N_g\times N_g$ number of patches, i.e, $K_g\times N_g\times N_g= M_r \times N_r$. It is noted that the number of partitions and the ensuing sub-surfaces are decided upon beforehand \cite{partition3}, and each subgroup is capable of providing a desired throughput. In this scenario, $R^t_i$ can be in one of two states: ON or OFF. An incident signal's phase can be altered to a desired direction when an element is in the ON state; when it is in the OFF state, it cannot reflect \cite{partition3}. Here, we use one particular subgroup instead of using total RIS, and we also assume that a subgroup can serve a single request at a particular instant \cite{can_serv_single_user}. Our objective is to select the energy-efficient subgroups for information transfer.
 \vspace{-4mm}
\subsection{Computation of Throughput and Energy-efficiency}\label{ener_thr}
In this communication scenario, let there be $n_d$ number of devices that lie within the free zones. We assume that any two devices $u$ and $v$ want to communicate with each other and they lie within the free zones $z_1$ and $z_2$, respectively. Now, depending on the position of the obstacles, $u$ and $v$ can communicate in three different ways as follows: $\rm (i)$ directly, $u$ and $v$ can communicate directly if there is no obstacle in between them, i.e., there is a LoS link between them. $\rm (ii)$ via single reflection, and $\rm (iii)$ via double reflections. We suppose that the wireless link experiences both small-scale block fading and large-scale path loss effects. The direct channel from $u$ to $v$ exhibits small-scale fading and  their corresponding path loss factor is $\rho_L^{\frac{1}{2}}d^{-\frac{\alpha}{2}}_{uv}$, where $\rho_L$ is the path loss at one meter distance, $\alpha$ is the path loss exponent and $d_{uv}$ denote the distance between $u$ and $v$.

We also assume that each group $ R^t_i$ of an RIS $R_i$ consists of $N_g \times N_g$ number of elements. Let $\mathbf{h}_{uR^t_i}\in \mathbb{C}^{N_g\times 1}, \mathbf{h}_{R^t_iR^s_j} \in \mathbb{C}^{N_g\times N_g} $ and $\mathbf{h}_{R^t_i/R^s_jv}\in \mathbb{C}^{1\times N_g}$ be the channel matrix from $u$ to $R^t_i$, $R^t_i$ to $R^s_j$ and, $R^t_i/R^s_j$ to $v$, respectively. 
Note that, unlike the conventional RIS-based approach, the grouping-based approach does not involve a diagonal phase shift matrix of non-zero $N_g \times N_g$ elements \cite{partition3}. A common phase shift is applied on the incoming signals, and the product of each point-to-point link determines the total path loss for each of these channel matrices \cite{2hop}. As a result, the effective channel gain for single reflection and double reflection is given by 
$\mathbf{h}_{R^t_iv} \mathbf{h}_{uR^t_i}\times e^{j\phi_{i,t}}$ and  $\mathbf{h}_{R^s_j} \mathbf{h}_{R^t_iR^s_j}\mathbf{h}_{uR^t_i}\times e^{j(\phi_{i,t}+\phi_{j,s})}$, where $e^{j\phi_{i,t}}$ and $ e^{j\phi_{j,s}}$ are the common phase shift of $R^t_i$ and $R^s_j$, respectively.

Now, we define a metric, data rate that will be used to quantify the performance of our proposed strategy. Here, we assume orthogonal frequency-division multiplexing (OFDM) \cite{freq_can} is used in our communication scenario. Let $P$ be the transmitted power, then the signal-to-noise-ratio (SNR) at the receiver for single RIS-assisted communication is 

\begin{equation}
    \gamma_{\rm sr}=\frac{P\rho_{\rm L}^2d_{R^t_iv}^{-\alpha}d_{uR^t_i}^{-\alpha}}{{\sigma^2}}\Big |\mathbf{h}_{R^t_iv}\times \mathbf{h}_{uR^t_i}\times e^{j\phi_{i,t}}\Big|^2,\label{sr}
\end{equation}
and SNR for double RISs reflected communication is 
\begin{equation}
     \gamma_{\rm dr}\!\!=\!\!\frac{P\rho_{\rm L}^3d_{R^s_jv}^{-\alpha}d_{uR^t_i}^{-\alpha}d_{R^t_iR^s_j}^{-\alpha}}{ \sigma^2}\!\Big |\mathbf{h}_{R^s_j} \times \mathbf{h}_{R^t_iR^s_j} \times \mathbf{h}_{uR^t_i}\times e^{j(\phi_{i,t}+\phi_{j,s})}\!\Big|^2\!,\label{dr}
\end{equation}
where $\sigma^2$ is the variance of the circularly symmetric zero mean additive white Gaussian noise. Note that, we adjust the common phase shift of \eqref{sr} and \eqref{dr} to attain the optimal SNR. Therefore, from Shannon's capacity formula, we can get the throughput
$T(\gamma)=\log(1+\gamma_{\rm{sr/dr}})$. Hence, by calculating the throughput, we can compare the performances.

Now, we define the energy-efficiency metric $(\rm E_{eff})$ \cite{dramp}, which will be used to measure how well our suggested approach performs. Let $T(\gamma)$ be the throughput obtained at the receiver end. Additionally, it requires a total $\rm E^i_c$ amount of energy for information transfer via a single RIS $R_i$, where
\vspace{-1mm}
\begin{equation}\label{ec9}
    {\rm E^i_c} = \beta \times \phi \times \frac{1}{T_i(\gamma)} \times \Big(P+P_{\rm {phase}}(R_i)\Big).
\end{equation}
For two consecutive RIS ($R_i$ and $R_j$) assisted communication the total energy consumption is denoted by $\rm E^{i,j}_c$ where
\vspace{-2mm}
\begin{equation}\label{ec10}
    {\rm E^{i,j}_c} = \frac{\beta  \phi}{T_{i,j}(\gamma)} \Big(P+P_{\rm phase}(R_i)+P_{\rm phase}(R_j)\Big),
\end{equation}
$\beta$ is the number of packets each with $\phi$ bits, $P$ is the transmitted power, and  $P_{\rm phase}(R_i)$ is phase shift power for RIS $R_i$.

Therefore, the energy-efficiency for single reflection and double reflections are defined by ${\rm E^i_{ eff}}=\frac{T_i(\gamma)}{\rm E^i_c} \quad\text{and}\quad
    {\rm E^{i,j}_{eff}}=\frac{T_{i,j}(\gamma)}{\rm E^{i,j}_c}$, respectively.
Moreover, we use this metric to select an appropriate subgroup for the information transfer of each blind pair.
\section{Problem Formulation}\label{prob}
Our objective is to place the minimum number of RISs in strategic locations and for each information transfer, we aim to find an energy-efficient subgroup for abstracted device pair. Therefore, we formulate the problem and break it down into two separate parts: i) RIS deployment and ii) Group selection. Now we are describing these two parts in detail below:
\vspace{-2mm}
\subsection{RIS Deployment}
In the first part, we want to formulate an optimization problem to cover a maximum number of blind pairs using the least number of RISs. It is noted that we do not consider those device pairs that have direct LoS and are totally blind. Here, let $\mathcal{B}$ be the set of all coverable blind pairs, i.e., no element of $\mathcal{B}$ remains uncovered after deploying the RISs. Moreover, to formulate the optimization problem, we define a few notations below.

Let $i$ and $j$ be two locations and
$a_{ij}$ be an indicating variable that describes the status of having LoS between them, where $1 \leq i,j \leq n^2$. That is,

\vspace{-6mm}
\begin{align}  \label{prop1211}
a_{ij}=\begin{cases} 
1, & \text{dist}\;(i,j)\leq r \;\& \; \exists \;\; \text{LoS between i and j }\\
0, & \text{else}
\end{cases}
\end{align}
Additionally, let $S_i$ be the set of all blind pairs covered by an RIS at the $i$-th cell.
That is, 
\begin{equation}
    S_i=\Big\{(p,q) : \;\;\;a_{pi}+ a_{iq}=2\Big\}\quad \forall \;\; i.
\end{equation}
Note that a blind pair of $\mathcal{B}$ may be visible via a single or double reflections. Therefore, we also consider $D_{ij}$ as a set of all blind pairs that are coverable via double reflections using two RISs located at $i$-th and $j$-th cells, but not coverable by either $i$-th or $j$-th RIS via a single reflection. Hence, $D_{ij}$ can be represented as
\vspace{-2mm}
\begin{align*}
    D_{ij}=\nonumber\Big\{(p,q):& \quad (p,j)\in S_i \;\;\& \quad(i,q)\in S_j, \quad(p,q)\notin S_i, \\ & \quad(p,q)\notin S_j \Big\} \quad i < j.
    \vspace{-2mm}
\end{align*}

However, we define $Z_{ij}$ as a set of all blind pairs that are coverable via single as well as double reflections using RIS placed at $i$-th and $j$-th locations. That is,
\vspace{-2mm}
\begin{equation}
    Z_{ij}=S_i \cup S_j \cup D_{ij}, \;\;\; i < j.
    \vspace{-2mm}
\end{equation}
\vspace{-1mm}
We now introduce the following binary variable:
\vspace{-2mm}
\begin{align*}  
x_i=\begin{cases} 
1, & \text{i-th cell is selected for RIS deployment},\\
0, & \text{else.}
\end{cases}
\end{align*}
Hence, we formulate the optimization problem as follows:
\vspace{-2mm}
\begin{align}\label{opt28}
{\rm Minimize} \quad & \sum_{i=1}^{n^2}
x_i\\
\textrm{such that} \quad & \bigcup_{i,j :i \leq j} Z_{ij}x_ix_j=\mathcal{B} \;\; \forall \;\;i < j.\tag{8.a}
\end{align}
This integer program can be linearized by using an intermediate binary variable $y_{ij}$ as follows:
\vspace{-2mm}
\begin{align}\label{opt}
{\rm Minimize} \quad & \sum_{i=1}^{n^2}
x_i\\
\textrm{such that} \quad & \bigcup_{i,j :i \leq j} Z_{ij}y_{ij}=\mathcal{B},\tag{9.a} \\
  & y_{ij}\leq x_j \quad \forall \;\;i < j ,\tag{9.b}\label{con1}\\
  & y_{ij} \leq x_i  \quad \forall \;\;i < j, \tag{9.c}\label{con2}\\
  & y_{ij}\geq x_i+x_j-1  \quad \forall \;\;i < j.\tag{9.d} \label{con3}
\end{align}

Note that the above integer linear program (ILP) is nothing but a classical set cover problem which is a well-known NP-hard problem. Therefore, in Section \ref{rdep}, we present a greedy solution for the RIS deployment problem.

\subsection{Group selection}

In the second part, for group selection, we assume that RISs have already been deployed. Moreover, we also know which blind pair will be covered by which RISs. Note that, a blind pair may be covered by single as well as double reflections. Hence, for a blind pair, more than one subgroup may be available to complete the information transfer. However, we allow a single subgroup for information transfer because of the scenario of energy constraints. Therefore, our primary objective is to identify a specific energy-efficient subgroup for each blind pair.
Let us assume that there are $n_b$ number of blind pairs, and $R$ number of RISs are deployed in the surroundings, and each of them is subdivided into $l$ number of subgroups. Let $\rm G^s_i$ be the set of all subgroups that can provide an indirect LoS link to the $i$-th blind pair via single reflection. Hence, we can represent $\rm G^s_i$ as
\vspace{-2mm}
\begin{equation*}
     {\rm G^s_i}\!=\!\Big \{R^k_j : \!\text{$i$-th pair is visible via  $k$-th subgroup of RIS $R_j$\!}\Big \}.
\end{equation*}

Similarly, let $\rm G^d_i$ be the set of all subgroups that can provide an indirect LoS link to the $i$-th blind pair via double reflections. That is,
 \vspace{-2mm}
\begin{align*}
    {\rm G^d_i}= \left\{R^{p,q}_{l,m}\;\;:\right.& \text{$i$-th pair is visible via $p$-th subgroup of RIS}\\ &  R_l \;\&\; \text{$q$-th subgroup of RIS $R_m$}\Big\}.
\end{align*}
Hence, if a blind pair is visible by single reflection, our problem is to find a more energy-efficient subgroup $R^{k^*}_{j^*}$ from $\rm G^s_i$ such that

\vspace{-3mm}
\begin{equation*}
    {\rm E^{k^*}_{eff}}\big(j^*\big)= \argmax \Big({\rm E^{k}_{eff}}\big(j\big)\Big).
    \vspace{-2mm}
\end{equation*}

If a blind pair is visible via double reflections, we select $R^{p^*,q^*}_{l^*,m^*}$ as energy-efficient subgroups from $\rm G^d_i$ such that  
\vspace{-3mm}
\begin{equation*}
    {\rm E^{p^*,q^*}_{eff}}\big(l^*,m^*\big)=\argmax \Big({\rm E^{p,q}_{eff}}\big(l,m\big)\Big).
    \vspace{-2mm}
\end{equation*}

Note that if a blind pair is visible via single as well as double reflections, we will select the more energy-efficient case for them. In Section. \ref{gsel}, we have discussed the group selection strategy in detail.
\vspace{-3mm}
\section{Proposed Strategy}\label{prop}
Here we discuss the RIS deployment and group selection strategy by considering both single and double reflections. Accordingly, we divide this section into two parts: i) RIS deployment strategy and ii) Group selection criteria.
In the first part, we propose a greedy deployment strategy, and in the second part, we investigate an energy-efficient group selection criteria. 

\begin{algorithm}[t!]

    \KwIn{$L, z_i\in Z, (u,v)$}
    \KwOut{Visible or not}
    Join $uz_i$ and $z_iv$\\
    \uIf{$uz_i\|z_iv$ ${\rm intersect \:\: at \:\: least \:\: one\:\: line \:\: segment\:\: of \:\: L}$}{not visible via $z_i$}
    \Else{Visible via $z_i$}
    \caption{Visibility Algorithm}
    \label{visibility}
\end{algorithm}
\subsection{Proposed RIS Deployment Strategy}\label{rdep}
In our proposed RIS deployment strategy, we first identify which blind pairs are present in the surroundings. After identifying the blind pairs, we will find the candidate locations for RIS deployment, and finally select the candidate locations for final deployment. All these steps are now discussed in detail below.
\subsubsection{\textbf{Blind Pairs Identification}}
Let there be $n_o$ obstacles and $n_d$ device pairs in a region, whose locations are known. We assume that the sides of an obstacle are formed by line segments and  $L$ is the set of all line segments of $n_o$ obstacles. Additionally, we assume that a device pair $(u,v)$ can communicate with each other if they reside within $r$ distance. Let $Z$ be the set of all free cells. Algorithm \ref{visibility} finds whether a device pair $(u,v)$ is visible via a free cell $z_i \in Z$ or not. Here, we aim to find all the blind pairs in a region. To achieve this, we first find the set of all device pairs that may be obstructed by a particular obstacle. Finally, continuing this process for all obstacles, we get the set of all blind pairs, which is described below.

\begin{algorithm}[t!]

    \KwIn{$L_i,\;\;D$}
    \KwOut{$B$}
    Initialize: $B= \phi$;\\
    \For{$i \in \{1, \cdots, n_o\}$}{
    $B_i=\phi, \; D_i= \phi$;\\
    \For{$(u^t,v^t)$ in $D$}{
    \If{$x^{\min}_i-r \leq u^t_x, v^t_x\leq x^{\max}_i+r\;\; \text{and} \;\;y^{\min}_i-r \leq u^t_y, v^t_y\leq y^{\max}_i+r$}{$D_i=D_i \cup \{(u^t,v^t)\}$}
    }
    \For{$(u^t,v^t)$ in $D_i$}{
    \For{$l \in L_i$}{
    \If{$l$ {\rm intersect} $u^tv^t$}{
    $B_i=B_i\cup \{(u^t,v^t)\}$\\
    Break;
    }
    
    }
    
    }
    $B=B\cup B_i$
    }
    Return $B$
    
    \caption{Blind Pairs Identification Algorithm}
    \label{bp}
\end{algorithm}

Let $O_i$ be the $i$-th obstacle, $L_i$ be the set of all line segments that constitute $O_i$, and 
\vspace{-2mm}
\begin{equation}
    C_i=\Big\{(x_t,y_t):\;\;\; t=1,\cdots n_i\Big\}
    \vspace{-1mm}
\end{equation}
be the set of all $n_i$ vertices of $O_i$. 
Therefore, let 
\vspace{-3mm}

\begin{align*}
    & x^{\max}_i=\max \big\{x_t:\;\;\; t=1,\cdots n_i\big\}, \nonumber \\ & x^{\min}_i=\min \big\{x_t:\;\;\; t=1,\cdots n_i\big\} \: \text{and}\\
    & y^{\max}_i=\max \big\{y_t:\;\;\; t=1,\cdots n_i\big\}, \quad  \nonumber \\ & y^{\min}_i=\min \big\{y_t:\;\;\; t=1,\cdots n_i\big\}
    \vspace{-2mm}
\end{align*}
be the maximum and minimum $x$-coordinate and $y$-coordinate of the vertices of $O_i$, respectively. Let
\vspace{-2mm}
\begin{align*}
    D &= \Big\{(u^t,v^t)\; : \;\;\; t=1,\cdots,n_d\Big\}\nonumber \\
    &=\Big\{ ((u^t_{x},u^t_{y}),(v^t_{x},v^t_{y})):\;\;\; t=1,\cdots,n_d\Big\}
    \vspace{-3mm}
\end{align*}
be the set of $x$ and $y$ coordinates of all the device pairs.
Let $D_i$ be the set of device pairs that could potentially be obstructed by $O_i$. That is,
\vspace{-2mm}
\begin{align}
    D_i=\Big\{(u^t,v^t) & \in D \;\; :\;\; x^{\min}_i-r \leq u^t_x,v^t_x \leq x^{\max}_i+r,\nonumber \\ &\;\; y^{\min}_i-r \leq u^t_y,v^t_y \leq y^{\max}_i+r \Big\}.
    \vspace{-2mm}
\end{align}
 Let $B_i$ be a set of all blind pairs that could potentially be obstructed by $O_i$. That is, 
\vspace{-2mm}
\begin{align}
    B_i=\left \{ (u^t,v^t) \in D_i\;\;: \right. & \quad \text{$u^tv^t$ intersect at least} \nonumber \\ & \left. \text{one line segment of}\; L_i \right \}.
    \vspace{-2mm}
\end{align}
Continuing this process, for each obstacle, we can compute the set $B$ of all blind pairs as follows: 
 \vspace{-3mm}
\begin{equation}
    B=B_1 \cup B_2 \cup \cdots \cup B_{n_o}=  \bigcup\limits_{i=1}^{n_o} B_i.
    \vspace{-2mm}
\end{equation}
The complete process of blind pair identification is shown in Algorithm \ref{bp}. Note that our main motivation is to serve a maximum number of blind pairs using indirect LoS via single or double reflections. In this context, we use a novel technique to find out the candidate zones for RIS deployment below.

\subsubsection{\textbf{Finding Candidate Locations for RIS Deployment}}\label{candid}
Our goal is to find the candidate locations for RIS deployment such that we can serve a maximum number of blind pairs with fewer RISs. Since the grid consists of $\mathbf{M}$ rows and $\mathbf{N}$ columns, there are $\mathbf{M}\times \mathbf{N}$ zones in the grid. Out of $\mathbf{M} \mathbf{N}$ zones, few are covered by obstacles, and the remaining are obstacle-free zones. Let there be $p$ free zones where $p< \mathbf{M} \mathbf{N}$ and $Z$ be the set of all free zones. That is, 
\vspace{-2mm}
\begin{equation}
    Z=\Big\{z_i:i=1,2,\cdots,p\Big\},
    \vspace{-2mm}
\end{equation}
where $z_i$ denotes the $i$-th free zone.

Let $A_i$ be the set of all blind pairs that are {\it visible} via $z_i$, i.e., for blind pair $(u,v) \in A_i$ there is a direct LoS between $u$ to $z_i$ and $z_i$ to $v$. Additionally, we denote the cardinality of $A_i$ as $card(A_i)$. Let $A$ be an array of $p$ elements whose $i$-th element represents the cardinality of $A_i$, i.e., $card(A_i)$. Let $\max (A)$ be the maximum element of $A$, and $t$ be the corresponding index of $\max (A)$ in the array $A$. Therefore, $z_t$ is the corresponding zone from where the maximum number of blind pairs can be served. Hence, we select $z_t$ as the first candidate location for RIS deployment and $A_t$ be the set of all blind pairs that are {\it visible} via $z_t$. Here we rename the set $A_t$ as $A^m_1$ and $z_t$ as $z_c^1$. Therefore, after finding the initial RIS location, let $B^1_{rem}$ represent the set of all the remaining blind pairs, i.e.,
\vspace{-3mm}
\begin{equation}
    B^1_{rem} =B \setminus A^m_1 = \big (A_1^m\big )^c,
    \vspace{-2mm}
\end{equation}
where $X^c$ denotes the complement of the set $X$.

\begin{figure}
    \centering
    \includegraphics[width=.8\linewidth]{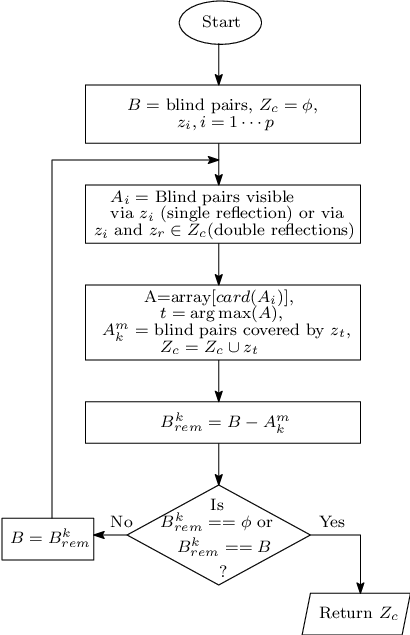}
    \caption{Proposed strategy for finding candidate locations}
    \label{algo2}
    \vspace{-6mm}
\end{figure}

After fixing the first candidate location $z_c^1$, the remaining blind pairs may be visible via i) single reflection or ii) double reflections. In particular, the remaining blind pairs in $B^1_{rem}$ may be visible via single reflection using any free cell other than $z_c^1$, or via double reflections using $z_c^1$ plus one of the remaining free cells. Therefore, we select that free cell as a second candidate location for RIS deployment from where the maximum number of elements of $B^1_{rem}$ will be served using single or double reflections. Let us denote $z_c^2$ as the second candidate zone for RIS deployment, and $B^2_{rem}$ as the set of remaining blind pairs. In a similar way,  we can find out the other candidate locations. Therefore, after finding the $k$-th candidate location,  the remaining  set of blind  pairs is given by
\vspace{-3mm}
\begin{equation}
    B^{k}_{rem}=\Big(A^m_1 \cup A^m_2 \cdots \cup A^m_{k-1} \cup A^m_k \Big)^c=B^{k-1}_{rem} \setminus A^m_k
    \vspace{-2mm}
\end{equation}
where $A^m_k$ is the set of all blind pairs that are served by the $k$-th candidate location, and $B^{k-1}_{rem}$ is the set of all remaining blind pairs before finding the $k$-th candidate location and the process will stop if $B^{k}_{rem}=B^{k-1}_{rem}$ or $B^{k}_{rem}=\phi$. Finally, we receive the set $Z_c$ of all candidate zones from where we can cover the greatest number of blind pairs. The RISs will actually be placed at the center of each free zone $z_i \in Z_c$. The proposed strategy is described in Fig. \ref{algo2}. 
\subsubsection*{{\bf Complexity and approximation ratio of the proposed algorithm}} Let there be $p$ free cells and $|B|$ many blind pairs. According to our proposed strategy, for selecting the first RIS location, we consider each free cell and compute the number of blind pairs that can be served by it using single reflection only. This requires $p|B|$ complexity. Next, for finding the k-th RIS location ($k \geq 2$), we consider each of the remaining $(p-k+1)$ free cells and compute how many blind pairs can be covered by it using single as well as double reflections. This requires $(p-k+1)|B| + (k-1)(p-k+1)|B| +  {k-1\choose 2} |B|$ complexity. Hence, the worst-case complexity of the proposed algorithm is $O(p^2|B|)$. This greedy algorithm is indeed the classic set cover greedy algorithm, which has a well-known approximate ratio of $O(\log n)$ \cite{vazirani}. The approximation ratio of our suggested RIS deployment algorithm is therefore $O(\log |B|)$, where $|B|$ represents the total number of blind pairs. In fact, it is also well-known that no polynomial-time algorithm for set cover can give a better approximation ratio than $O(\log n)$ unless P = NP \cite{vazirani}.

From the above discussion and observation, we get the following remarks.

\begin{rem}
      If $B^{k}_{rem} \neq \phi$ and $B^{k}_{rem}=B^{k-1}_{rem}$ holds then $B_{un}=B^{k}_{rem}$ is the total number of uncovered blind pairs. That is,
        $B_{un}=\Big(A^m_1 \cup A^m_2 \cdots \cup A^m_{k-1} \cup A^m_k \Big)^c$.
\end{rem}
\begin{rem}
    If only $k$ number of RISs are allowed to deploy, then the total number of blind pairs that can be served is  given by
    $\Big(B^{1}_{rem} \cap B^{2}_{rem} \cap \cdots \cap B^{k}_{rem}\Big)^c=B \setminus B^{k}_{rem}$.
    
\end{rem}

\begin{rem}
    In a particular scenario, if no blind pair is visible via double reflections, then our proposed strategy will be converted into the RIS deployment strategy for single reflection.
\end{rem}

Below, we have illustrated the proposed RIS deployment strategy using an example.

\subsubsection{\textbf{Illustrative Example of the Proposed Deployment Strategy}}
\begin{figure}
    \centering
    \includegraphics[width=.32\linewidth]{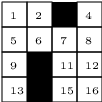}
    \caption{Example of a communication environment}
    \vspace{-6mm}
    \label{fig_grid}
\end{figure}

A specific scenario of our proposed strategy is demonstrated in Fig. \ref{fig_grid}. Here we consider a grid that consists of four rows and four columns. As this grid consists of sixteen cells, we label these cells from $1$ to $16$. Moreover, we assume that the obstacles are located in the black cells, which correspond to the cell numbers $3,10$, and $14$. We also consider that a device can lie only within a free cell. Here, $\{1,2,4,5,6,7,8,9,11,12,13,15,16\}$ is the set of all free cells. Here, we assume that each pair of cells is within $r$ distance apart from each other. Therefore, in this communication environment, 
\vspace{-6mm}

\vspace{-.5mm}
\begin{align*}
    B\!=\!\{&(1,4),(1,8),(1,11),(1,15),(1,16),(2,4),(2,7),
    (2,8),\\
    &\!\!\!\!\!\!(2,13),(2,12),(2,15),(4,5),(4,6),(4,7),(4,9),(4,13),\\
    &\!\!\!\!\!\!(5,12),(5,11),(5,15),(5,16),(6,9),(6,11),(6,13),\\
    &\!\!\!\!\!\!(6,15),(6,16),(7,9),(7,13),(8,9),(8,13),(9,11),(9,12),\\
    &\!\!\!\!\!\!(9,15),(9,16),(11,13),(12,13),(13,15),(13,16)\}
\end{align*}
is the complete set of all the blind pairs. Specifically, there is no direct LoS link between any device pair of $B$. As obstacles block the direct LoS of the blind pairs, RISs can be strategically deployed to provide an indirect LoS link. It can be observed that each of the blind pairs in $\big\{(4,9),(4,13),(9,15),(11,13),(12,13),(13,15),(13,16)\big\}$ can not be covered via single reflection, even if we deploy RISs in all the free cells.
However, in our proposed strategy, we can serve more blind pairs using double reflections along with the single reflection. It can be observed that the maximum number of blind pairs can be served by placing an RIS at cell $12$. Thus, our strategy will choose cell $12$ as the first RIS location. Also, note that cells $1$ and $12$ together can serve all the remaining blind pairs using single and double reflections. That is, according to our proposed strategy, $\big\{1,12\big\}$ will be the set of locations for RIS deployment.

In the following Subsection \ref{gsel}, we will address which subgroup of an RIS will be selected for information transformation.

\subsection{Group Selection Criteria}\label{gsel}
In section \ref{rdep} above, we have strategically found $R$, the set of all deployed RISs, and the set of all coverable blind pairs $\mathcal{B}=B\setminus B_{un}$. A blind pair $(u,v) \in \mathcal{B}$ can be covered in three different ways: i) via only single reflection, ii) via only double reflections, and iii) via both single and double reflections. Since energy consumption is a very important parameter in wireless communication scenarios, we discuss in the following the process to find the energy-efficient group considering the above-mentioned three cases. 

\begin{figure}[t!]
    \centering
    \includegraphics[width=.9\linewidth]{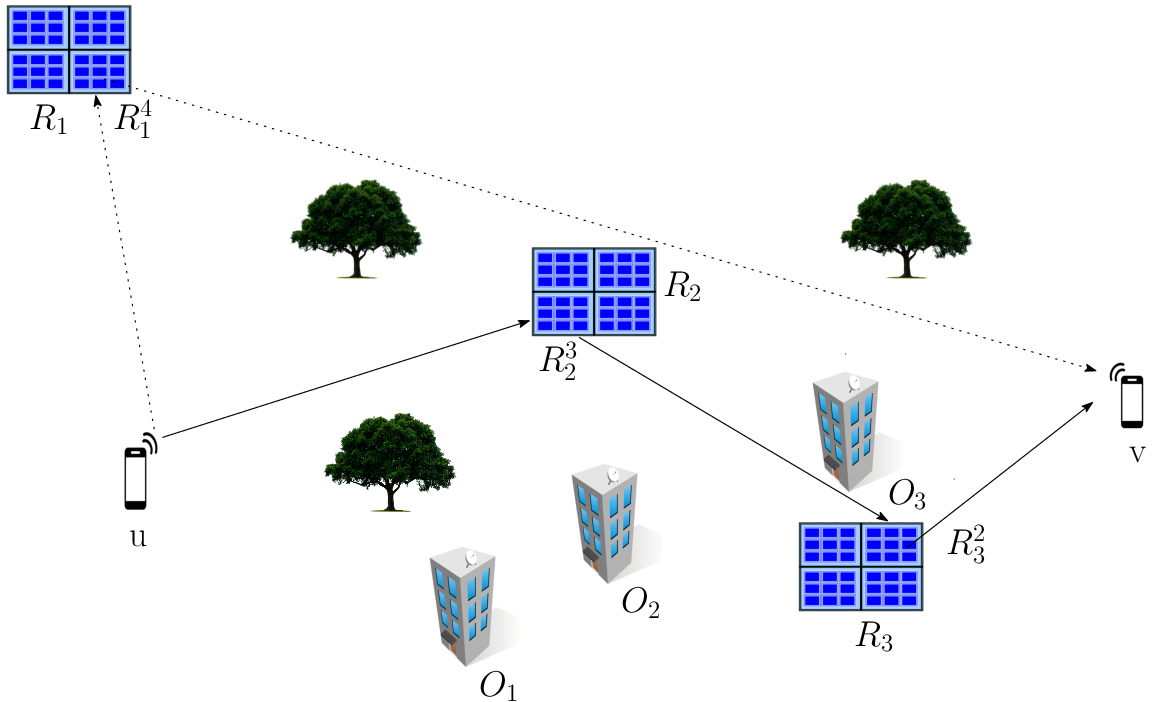}
    \caption{Sometimes double reflections are more energy-efficient than single reflection}
    \label{group}
    \vspace{-6mm}
\end{figure}

\subsubsection{\textbf{Energy-efficient Subgroup Selection for single reflection}}\label{single}
In this case, $u$ sends all the packets to $v$ using one subgroup of an RIS as an intermediate reflector. Let $S_{u,v}$ be the set of RISs each of which can provide indirect LoS between $u$ and $v$ via single reflection. That is, $S_{u,v}=\{R_i: \;\;R_i\;\text{is visible from both} \;$u$ \;\text{and}\; v\}$.
 Since each RIS is partitioned into $k$ non-overlapping subgroups, different subgroups may provide different data rates. In this context,  from subsection \ref{ener_thr},
the required energy $\rm E_c(i)$ for information transfer from $u$ to $v$ using $i$-th subgroup of $s$-th RIS is given by
\vspace{-3mm}
\begin{equation}\label{ec}
   {\rm E^s_c}(i) = \beta \times \phi \times \frac{1}{T^i_s(\gamma)} \times \Big(P+P_{\rm phase}(R^i_s)\Big),
    \vspace{-2mm}
\end{equation}
where $ P_{\rm phase}(R^i_s)$ is the phase shift power of $i$-th subgroup of $R_s$.

Therefore, the energy-efficiency ${\rm E^s_{eff}}(i)$ for a particular $(u,v)$ pair using $i$-th subgroup is 
\vspace{-2mm}
 \begin{equation}\label{ef5}
     {\rm E^s_{eff}}(i)=\frac{T^i_s(\gamma)}{{\rm E^s_c}(i)}.
     \vspace{-2mm}
 \end{equation}
 Now, we form an optimization problem to select an energy-efficient subgroup as below:
 \vspace{-2mm}
 
\begin{align}\label{opt1}
\max \quad & {\rm E^s_{eff}}(i)\\
\textrm{s.t.} \quad &  T^i_s(\gamma) \geq T_{th}, \;i=1, \cdots k \;\; \text{and} \;\; s \in S_{u,v},\tag{19.a}\\
  & {\rm E^s_c}(i) > 0 \;\;\;\;i=1, \cdots k\;\; \text{and} \;\; s \in S_{u,v}\tag{19.b},
  \vspace{-3mm}
\end{align}
where $T_{th}$ is a predefined threshold.

\begin{algorithm}[h]

    \KwIn{$R, (u,v)$}
    \KwOut{$R^{i^*}_{s^*}, R^{l^*,m^*}_{i^*,j^*}$}
    \CommentSty{\% Single Reflection} \nonumber \\
    \For{$s \in R$}{
    \uIf{$(u,v)$ visible via $R_s$}{
    \For{$i \in R_s$}{compute ${\rm E^s_{eff}}(i)$}
    
    ${\rm E^{s^*}_{eff}}\big(i^*\big)= \argmax \Big({\rm E^{s}_{eff}}\big(i\big)\Big)$}
    \Else{${\rm E^{s^*}_{eff}}\big(i^*\big)=0$}}
    \CommentSty{\% Double Reflection}\\
    \For{$i \in R$}{
    \For{$j \in R$}{
    \uIf{$(u,v)$ visible via $R_i$ and $R_j$}{
    \For{$l \in R_i$}{
    \For{$m \in R_j$}{
    \If{$(u,v)$ visible via $R^l_i$ and $R^m_j$}{Compute $ {\rm E^{i,j}_{eff}}{\big(l,m\big)}$}
    }
    }
    ${\rm E^{i^*,j^*}_{eff}}\big(l^*,m^*\big)=\argmax \Big({\rm E^{i,j}_{eff}}\big(l,m\big)\Big)$
    }
    \Else{${\rm E^{i^*,j^*}_{eff}}\big(l^*,m^*\big)=0$}
    }
    }
    \CommentSty{\% Both Single and Double Reflection}\\
    \uIf{${\rm E^{s^*}_{eff}}\big(i^*\big) \geq {\rm E^{i^*,j^*}_{eff}}\big(l^*,m^*\big)$}{Return $R^{i^*}_{s^*}$}
    \Else{Return $R^{l^*,m^*}_{i^*,j^*}$}
    \caption{Group Selection Algorithm}
    \label{gs_algo}
   \end{algorithm}
\begin{figure*}[h!]
 \begin{subfigure}[b]{.32\textwidth}
    \centering
    \includegraphics[width=0.92\linewidth]{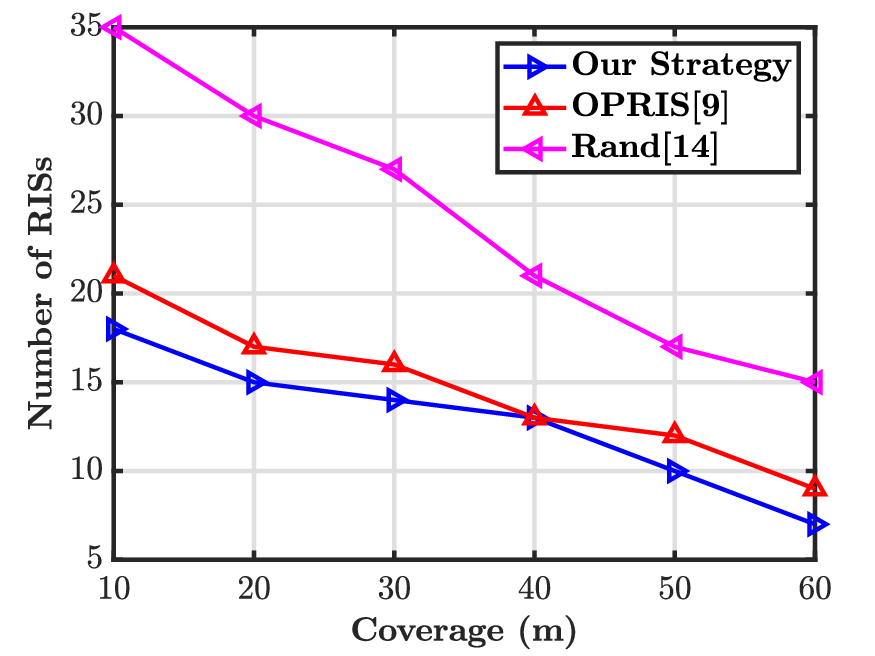}
    \vspace{-2mm}
    \caption{}
    \vspace{-2mm}
    \label{cov}
\end{subfigure}
\begin{subfigure}[b]{.32\textwidth}
    \centering
    \includegraphics[width=0.92\linewidth]{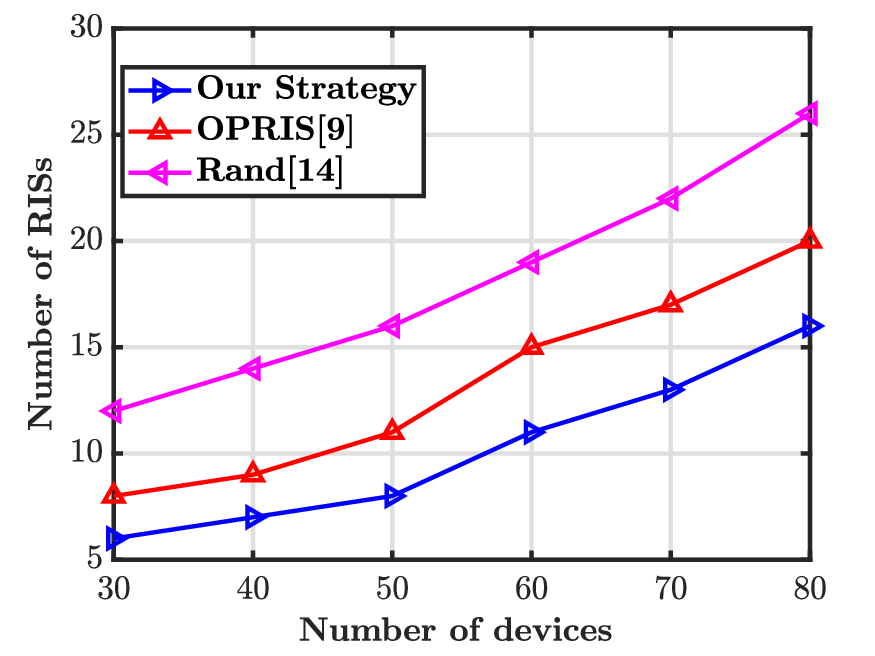}
    \vspace{-2mm}
    \caption{}
    \vspace{-2mm}
    \label{div}
\end{subfigure}
\begin{subfigure}[b]{.32\textwidth}
    \centering
    \includegraphics[width=0.92\linewidth]{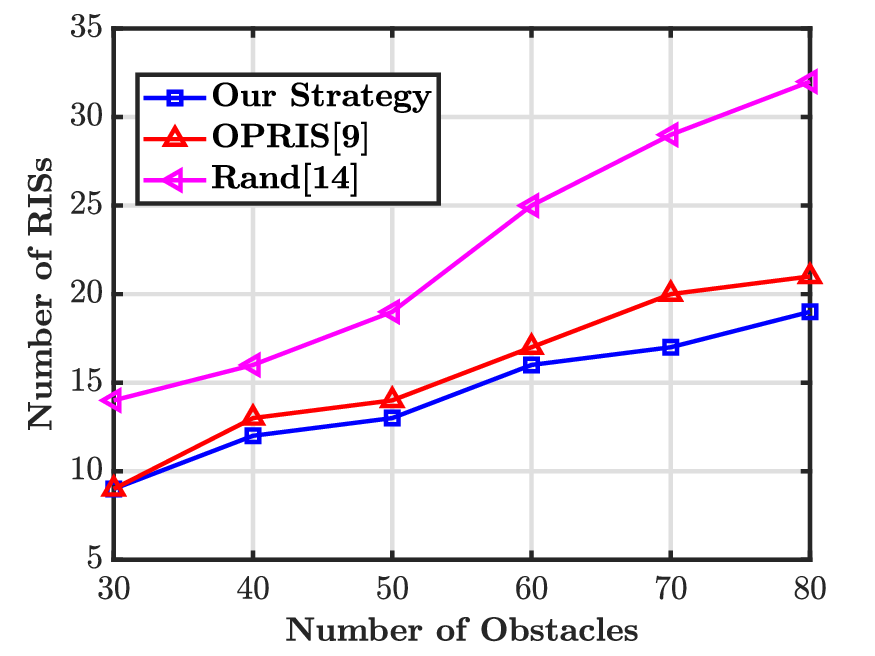}
    \vspace{-2mm}
     \caption{}
    \vspace{-2mm}
    \label{obs}
\end{subfigure}
\caption{\footnotesize  Impact of (a) Coverage area (b) Number of devices, and (c) Number of obstacles on number of RISs.}
\vspace{-4mm}
\end{figure*}
\subsubsection{\textbf{Energy-efficient Subgroup Selection for Double Reflections}}\label{double}
In this communication environment, $u$ transfers the packets to $v$ using two consecutive RISs $R_i$ and $R_j$, respectively. We also assume that $D_{u,v}$ be the set of RISs each of which can provide indirect LoS between $u$ and $v$. That is, $D_{u,v}=\{(R_i,R_j): \;\;\; (R_i\;\text{is visible from  $u$} \;\text{and}\; R_j)\; \& \;(R_j \;\text{is visible from}\;v)\}$. Since each RIS is partitioned into $k$ non-overlapping subgroups, one subgroup from $R_i$ and another from $R_j$ will be selected for complete information transfer from $u$ to $v$. Therefore, from \eqref{ec10}, the required energy for complete information transfer using $l$-th and $m$-th subgroup of $R_i$ and $R_j$ respectively, is given by
\vspace{-2mm}
\begin{equation}\label{ec1}
   {{\rm E^{i,j}_c}(l,m)} = \frac{\beta  \phi}{T^{l,m}_{i,j}(\gamma)} \Big(P+P_{\rm phase}(R^l_i)+P_{\rm phase}(R^m_j)\Big),
    \vspace{-2mm}
\end{equation}
where, $P_{\rm phase}(R^m_j)$ and $P_{\rm phase}(R^l_i)$ are the phase shift power for $m$-th and $l$-th subgroup of $R_j$ and $R_i$, respectively.
Moreover, we can compute the energy-efficiency as 
\vspace{-2mm}
\begin{equation}\label{ef1}
     {{\rm E^{i,j}_{eff}}(l,m)}=\frac{T^{l,m}_{i,j}(\gamma)}{ {\rm E^{i,j}_c}(l,m)}.
     \vspace{-2mm}
 \end{equation}

 Now, we formulate an optimization problem to select the energy-efficient subgroups as follows: 
 \vspace{-4mm}
 
 \begin{align}\label{opt2}
\max \quad & {\rm E^{i,j}_{eff}}(l,m)\\
\textrm{s.t.} \quad & T^{l,m}_{i,j}(\gamma) \geq T_{th}, \;l,m=1, \cdots k \;\&\; i,j \in D_{u,v} \tag{22.a}\\
  & {\rm E^{i,j}_c}(l,m) > 0 \;\;\;\;l,m=1, \cdots k\;\&\; i,j \in D_{u,v}.\tag{22.b}
  \vspace{-2mm}
\end{align}

\subsubsection{\textbf{Energy-efficient Subgroup Selection for both Single and Double Reflections}}
In this communication environment, a device pair $(u,v)$ is covered by both single and double reflections. That is, $S_{u,v}\neq \phi$ and $D_{u,v}\neq \phi$. Therefore, using case \ref{single} above, we get a subgroup that provides the maximum energy-efficiency for single reflection. Similarly, using case \ref{double} above, we get a pair of subgroups that provide the maximum energy-efficiency for double reflections. Finally, we select the best among them for information transfer between $u$ and $v$. Note that once the RISs are deployed, the time required for establishing the connection for a particular requesting pair $(u,v)$ is $O(|Z_c|)$, where $|Z_c|$ denotes the total number of RISs deployed.

In the following propositions, we will prove two interesting facts for a device pair covered via both single and double reflections.

\begin{proposition}\label{prop1}
    Let a device pair $(u,v)$ be visible through $R^l_i$ using single reflection, and through $R^l_i$ and $R^m_j$ together via double reflections. In that case, single reflection are always more energy-efficient than double reflections. 
\end{proposition}
\begin{proof}
    See Appendix \ref{app2}.
\end{proof}
From \eqref{ec} and \eqref{ec1}, we observe that energy consumption is a function of data rate, transmit power, and phase shift power. Again, the data rate is a function of distance.  Moreover, if a device pair $(u,v)$ is visible via $R^t_s$ using single reflection, and $R^l_i$ and $R^m_j$ for double reflections where $R^t_s, R^l_i$ and $R^m_j$ are three distinct subgroups, the result may be quite different, which is counter-intuitive. More specifically, in the following proposition, we prove that double reflections may be more energy-efficient than single reflection depending on the power consumption and the distance traveled by the signal.

\begin{proposition}\label{d_better}
Let a device pair $(u,v)$ be visible through $R^t_s$ using single reflection, and through $R^l_i$ and $R^m_j$ together via double reflections, where $R^t_s, R^l_i$ and $R^m_j$ are three distinct subgroups. If the following conditions hold:
\vspace{-2mm}
\begin{align*}
    &{\rm i})\;\;  P+P_{\rm phase}(R^t_s) \leq P+P_{\rm phase}(R^l_i)+P_{\rm phase}(R^m_j) \;\; \text{and}\\
    &{\rm ii})\;\; \rho_{\rm L} d_{R^t_sv} d_{uR^t_s} \geq d_{R^m_jv} d_{uR^l_i} d_{R^l_iR^m_j}
    \vspace{-4mm}
\end{align*}
  then double reflections are more energy-efficient than single reflection.  
\end{proposition}

\begin{proof}
    See Appendix \ref{app3}.
\end{proof}

In the following subsection, we illustrate with an example, the situation where double reflections are more energy-efficient than single reflection.

\subsubsection{\textbf{Illustrative Example where Double Reflection is more Beneficial than Single Reflection}}
In Fig. \ref{group}, $u$ wants to communicate with $v$ and two possible paths are available in between them, one is via $R^4_1$ using single reflection, and another one via $R^3_2$ and $R^2_3$ together using double reflections. Since $R^4_1$ is located far away from $u$ and $v$, the achieved data rate at $v$ is not very high. Moreover, due to the close proximity of $u$ to $R^3_2$, $R^3_2$ to $R^2_3$, and $R^2_3$ to $v$ the achievable data rate at $v$ is more than the achievable data rate via the path using $R^4_1$, which follows from Proposition \ref{d_better}. Therefore, in this case, the second path is more energy-efficient. As a result, $u$ selects the path via $R^3_2$ and $R^2_3$ for complete information transfer using double reflections.

\vspace{-2mm}
\section{Simulation Results}\label{simu}
In this section, we conduct comprehensive simulations to verify the effectiveness of our suggested approach and compare it with the closest available methods \cite{opris}, \cite{ran} and \cite{dar}. Here we consider a two-dimensional square area of $400\times400 \; m^2$ \cite{deb2021ris}. Furthermore, we assume that two devices can communicate only if they lie within a specific coverage radius. Moreover, we anticipate that D2D communication will occur at a frequency of $60$ GHz with a $500$ MHz bandwidth \cite{bndwidth}. The transmit power $P$ is $30$ dBm \cite{dramp} and $P_{\rm phase}=5$ dBm \cite{energy_eff}. Here, we consider a Rician fading scenario \cite{tvt}, incorporating a Rician factor $K=10$ dB \cite{dramp}. The default parameters taken into account are: path loss at one-meter distance $\rho_L=10^{-3.53}$ \cite{prmtre1}, the path loss exponent is $\alpha=2$ and the packet length is $1000$ bits. Below we briefly describe the existing methods \cite{opris}, \cite{ran} and \cite{dar} with which we have compared our proposed approach.
\begin{itemize}
    \item OPRIS \cite{opris}: This work investigates the optimal placement of RISs for a single reflected scenario. In other words, it does not allow double reflections. 
    \item Rand \cite{ran}: In this work, the RISs are placed arbitrarily in the geographically separated locations. Here too, the authors investigate the aspect of single reflection. 
    \item DAR \cite{dar}: This work proposes an approach to connect a particular device pair by considering the aspect of double reflections. However, it does not include the feature of RIS grouping in the communication scenarios.
\end{itemize}

\begin{figure*}[t]
\begin{subfigure}[b]{0.5\textwidth}
    \centering
    \includegraphics[width=0.66\linewidth]{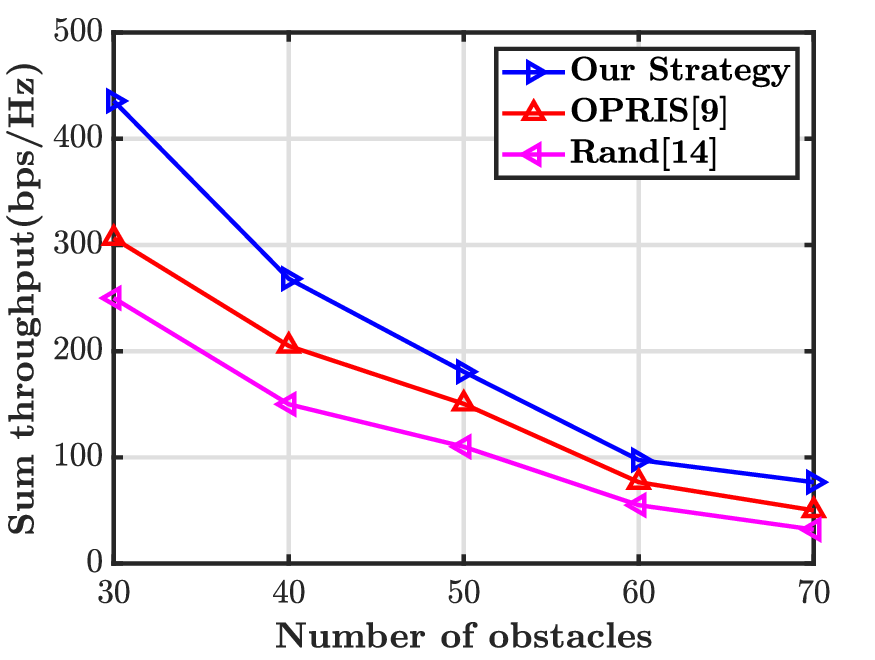}
    \caption{}
    \label{thu}
\end{subfigure}
\begin{subfigure}[b]{0.5\textwidth}
    \centering
    \includegraphics[width=0.66\linewidth]{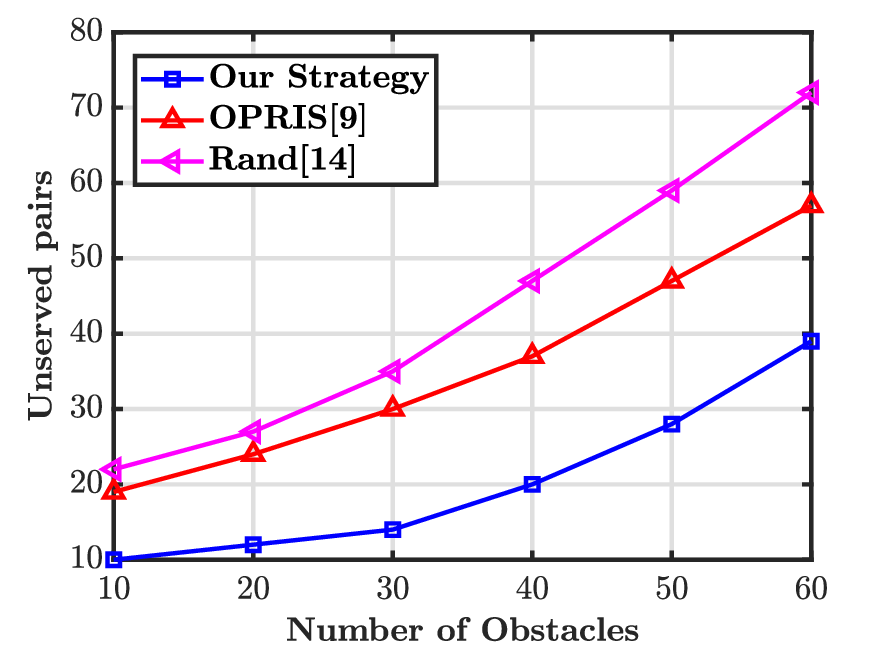}
     \caption{}
    \label{obs1}
\end{subfigure}
\vspace{-4mm}
\caption{\footnotesize  Impact of the number of obstacles on (a) Sum throughput, and (b) Unserved pairs.}
\vspace{-4mm}
\end{figure*}

\begin{figure*}[t]
 \begin{subfigure}[b]{.5\textwidth}
    \centering
    \includegraphics[width=0.66\linewidth]{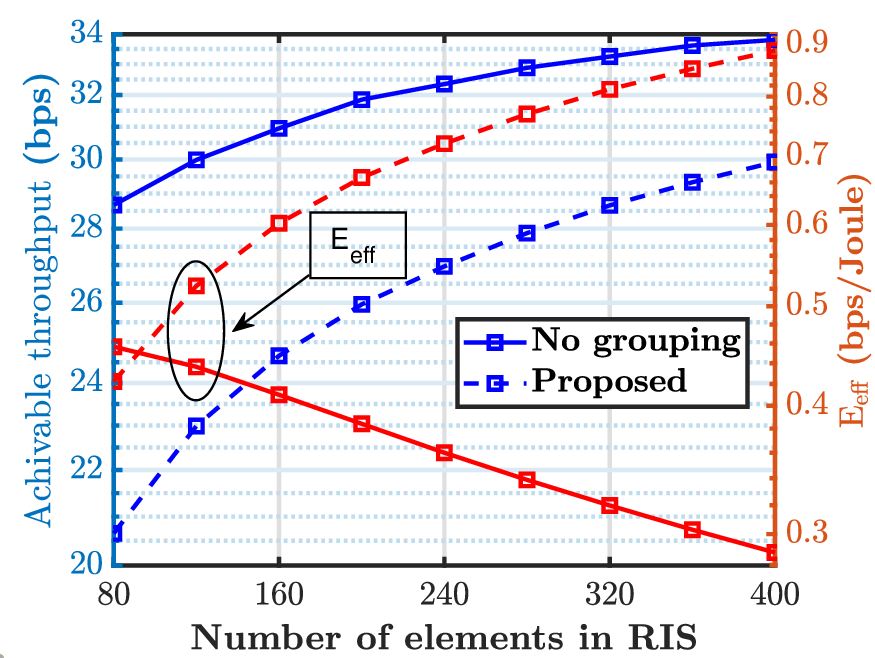}
    \caption{}
    \label{EC}
\end{subfigure}
\begin{subfigure}[b]{.5\textwidth}
    \centering
     \includegraphics[width=0.66\linewidth]{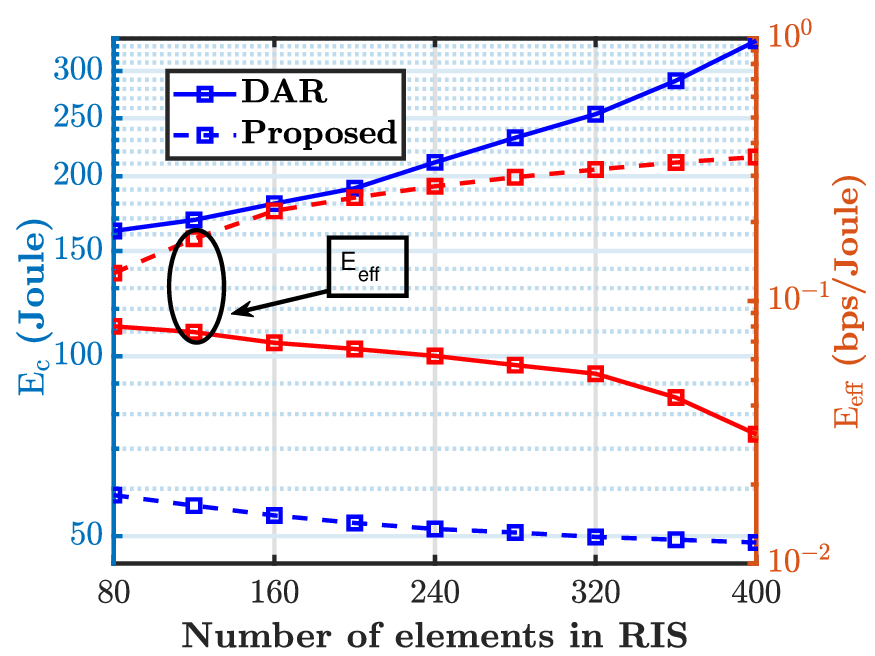}
    \caption{}
    \label{Ef}
\end{subfigure}
\caption{\footnotesize  Impact of the number of elements in a group on (a) energy consumption and (b) energy-efficiency.}
\vspace{-6mm}
\end{figure*}

Fig. \ref{cov} shows how many RISs are needed to cover the largest possible region as a function of coverage radius. Here, we consider three different scenarios with three different RIS placement strategies. Consequently, we look at the number of RISs used in these scenarios to get the maximum coverage. Note that, with increasing coverage radius, the number of RISs used exhibits a non-increasing trend, which is quite intuitive. We observe that our strategy outperforms both \texttt{OPRIS} and \texttt{Rand}. This is because, there are many device pairs which are visible via double reflections, but for a single reflection, we need to install a new RIS. As a result, our strategy brings down the requirement of RISs in comparison to \texttt{OPRIS} and \texttt{Rand}, as they did not allow double reflections.

Fig. \ref{div} and Fig. \ref{obs} show how many RISs are needed to cover the largest possible region as a function of device density and number of obstacles, respectively. Accordingly, we look at the number of RISs used in three different scenarios to get the maximum coverage. We find that, for a given situation, the number of RISs increases with the growing density of devices (obstacles), which is quite intuitive. Here too, we observed that our strategy outperforms \texttt{OPRIS} and \texttt{Rand} for the same reason as stated earlier.

\begin{figure*}[t]
\begin{subfigure}[b]{.32\textwidth}
    \centering
    \includegraphics[width=\linewidth]{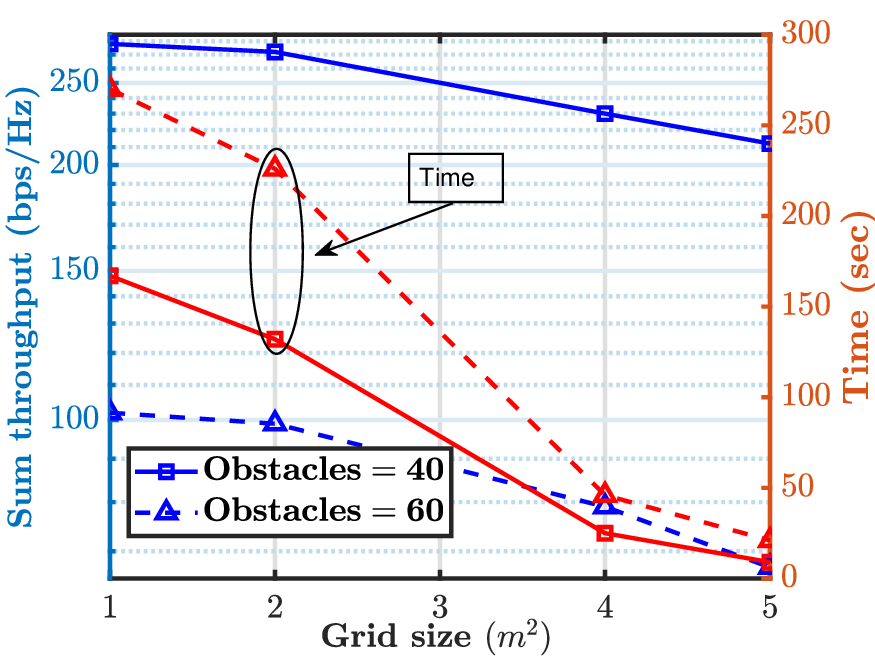}
    \caption{}
    \label{grid_resulation}
\end{subfigure}
\begin{subfigure}[b]{.32\textwidth}
    \centering
    \includegraphics[width=\linewidth]{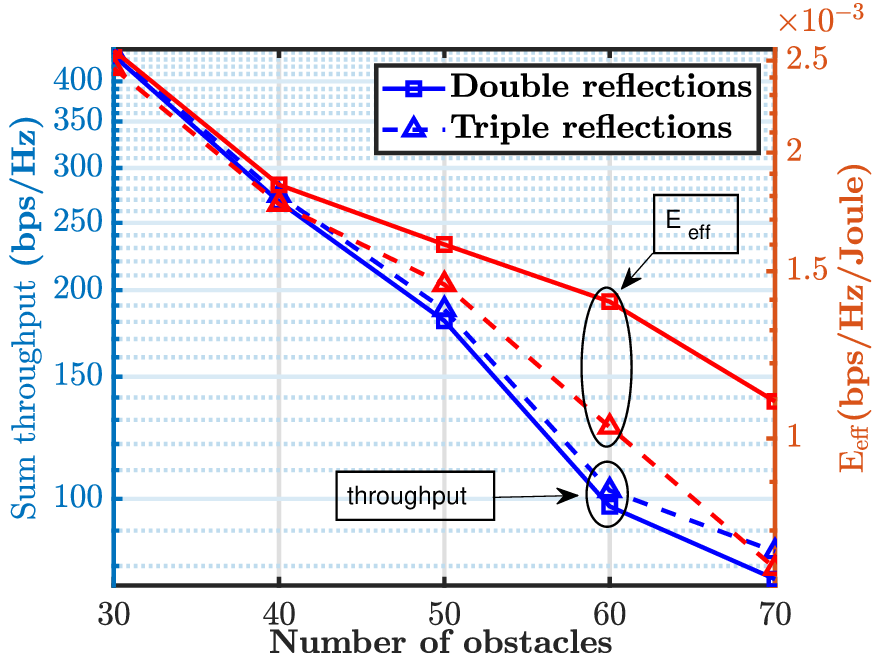}
    \caption{}
    \label{multi_double}
\end{subfigure}
\begin{subfigure}[b]{.32\textwidth}
    \centering
    \includegraphics[width=\linewidth]{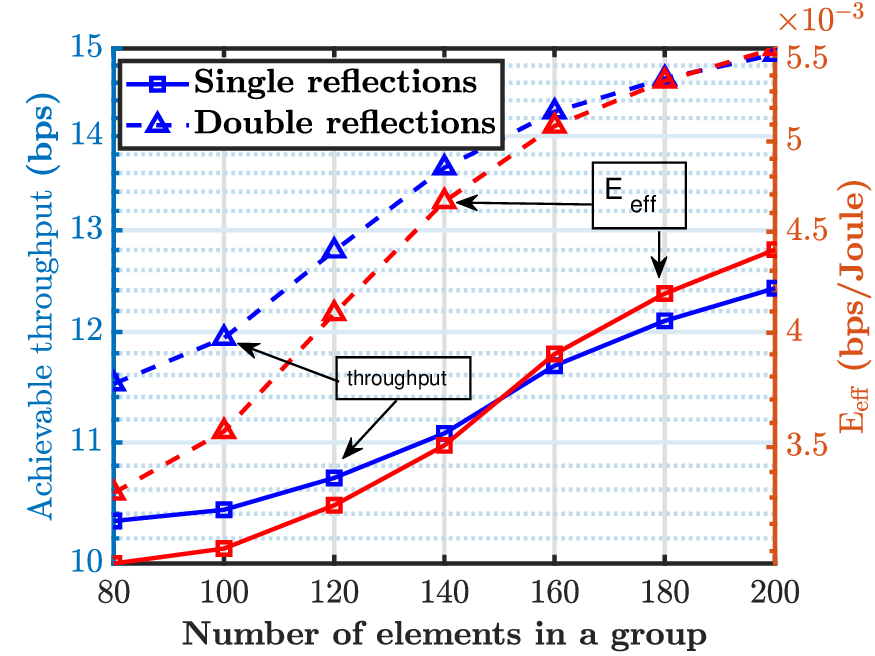}
     \caption{}
    \label{single_double}
\end{subfigure}
\vspace{-2mm}
\caption{\footnotesize  Performance trade-off investigation: (a) different obstacles, (b) double and triple reflections, and (c) Proposition $1$.}
\vspace{-6mm}
\end{figure*}

Fig. \ref{thu} demonstrates the impact of obstacle density on the sum throughput, where the sum throughput is calculated by using \eqref{sr} and \eqref{dr}. Here, we observe that, for a particular scenario, the sum throughput shows a decreasing trend with an increasing number of obstacles. Moreover, our proposed strategy outperforms \texttt{OPRIS} and \texttt{Rand} in terms of sum throughput. This is because in our proposed strategy, we can serve more blind pairs due to considering the double reflections.

From Fig. \ref{obs1}, we observe that, in relation to the growing obstacles, the number of unserved device pairs exhibits an increasing trend. As the number of obstacles increases, the number of blind pairs also increases. As a result, the likelihood of having more blind pairs that are not served rises. Note that using double reflections, sometimes we can serve some blind pairs that are not possible to serve by single reflection even after deploying more RISs.  As a result, our proposed strategy outperforms the performance of \texttt{OPRIS} and {\texttt{Rand} in terms of number of unserved blind pairs.

The significance of RIS grouping on the achievable throughput and energy-efficiency, as a function of the number of elements in a RIS, is shown in Fig. \ref{EC}. In particular, we examine the effects of distinguishing between a grouping-based scenario (GBS) and a non-grouping-based scenario (nGBS). In nGBS, the entire RIS is used for information transfer, whereas in GBS, the RIS is divided into four equal-sized groups, and only one of them is used for information transfer.
Here, we observe that in both cases, achievable throughput follows an increasing trend with the growing number of elements. This is because a growing number of patches support better throughput. As a result, nGBS provides higher data throughput as compared to GBS. This makes sense because, in contrast to GBS, nGBS makes use of the entire RIS, whereas GBS only uses a portion of the RIS's total number of patches.
We also observed from \ref{EC} that, in relation to the rising number of elements in a RIS, the $\rm E_{eff}$ for nGBS is in a decreasing trend whereas it is in an increasing trend for GBS. This is because the required phase shift power in nGBS is proportional to the number of patches of an RIS, whereas a common phase shift power is used for GBS. As a result, usage of the GBS leads to better $\rm E_{eff}$ performance.

 In Fig. \ref{Ef}, a comparison of energy consumption $(\rm E_c)$ and energy-efficiency $(\rm E_{eff})$ between the proposed strategy and the existing benchmark \texttt{DAR}, are shown. Here, as mentioned in \ref{EC}, we partitioned a RIS into four equal parts and chose one of them to facilitate communication. Additionally, we consider a scenario, where both strategies use double reflections.
 In \ref{Ef}, we observed that in relation to the growing number of patches of an RIS, $\rm E_{c}(\rm E_{eff})$ exhibits an increasing (decreasing) trend in \texttt{DAR}, and $\rm E_{c}(\rm E_{eff})$ exhibits a decreasing (increasing) trend in our proposed strategy. This improvement is because of the fact that the required phase shift power in \texttt{DAR} is proportional to the number of patches of an RIS as it uses the entire RIS, whereas a particular group is being used, and a common phase shift power is used in our proposed strategy. 

 Fig. \ref{grid_resulation} shows the impact of grid size on both sum throughput and computation time for different numbers of obstacles. Here, grid size $a$ signifies that the service area is discretized as $a \times a$ squares.  Here, we observe that as the grid size increases, both the sum throughput and the computation time exhibit a decreasing trend. As stated in the system model, we assumed that an obstacle occupies the entire grid, even if it just occupies a portion of it.
 Consequently, the total number of LoS links decreases significantly, leading to a downward trend in the sum throughput. On the other hand, as the size of the grid increases, the overall computation time decreases, as computation time is a function of the number of grids. Additionally, the performance of $40$ obstacles is superior to that of $60$ obstacles, which is also quite intuitive.

 Fig. \ref{multi_double} illustrates the importance of double and triple reflections on the achievable throughput and energy efficiency $\rm E_{eff}$ as a function of the number of obstacles. The sum throughput exhibits a decreasing trend with the number of obstacles for both double and triple reflections, which is quite intuitive.  It can be observed that the sum throughput for triple reflection is only marginally higher than double reflections. Moreover, this difference is more towards the higher number of obstacles, i.e., for the extremely obstructed environments. On the other hand, the energy-efficiency is not as good as that of double reflections. It can be observed that the improvement in sum throughput is significant when we take into account the second reflections. However, due to the inclusion of the third reflections, the improvement in sum throughput is found to be only marginal. Moreover, this marginal improvement in sum throughput is obtained at the cost of significantly more energy consumption. This, in turn, reduces the energy efficiency for triple reflections. This further justifies the widespread assumption of ignoring the triple and higher order reflections in the literature.

Comparisons of the achievable throughput and energy-efficiency for single and double reflections using our proposed approach are presented in Fig. \ref{single_double}. According to Proposition \ref{prop1}, double reflection is more energy-efficient than single reflection under certain conditions. In support of this, we consider a fixed $S-D$ pair that is coverable by both single and double reflections, and it satisfies the conditions of Proposition \ref{prop1}. Here, we observe that, in relation to the growing number of elements of a group, the achievable throughput and energy-efficiency for both single and double reflections exhibit an increasing trend and saturate after a certain number of elements of a group. Here, our proposed strategy for double reflection outperforms the performance of single reflection in terms of achievable throughput and energy-efficiency, which is quite counter-intuitive. The reason for this is the longer distance between $S$ and RIS, as well as RIS and $D$, for single reflection. Whereas, for double reflection, the $S-D$ pair is served by multiple short hops using two consecutive RISs.

\section{Conclusion}\label{con}
 In this work, we proposed a novel RIS deployment strategy in the double RIS-assisted D2D wireless communication scenario, which takes into account the elements grouping of an RIS. The proposed strategy prevents resource wastage by deploying the RIS strategically taking care of both single and double reflections and partitioning the RIS into non-overlapping subgroups. Subsequently, we proposed an energy-efficient group selection strategy for a device pair to complete their communication. It is interesting to note that under some conditions, double reflections are more energy-efficient than single reflection. The simulation results show that a significant reduction in the number of RISs is achievable by allowing double reflections. In addition, double reflections provide more energy efficient communication, and also bring down the number of unserved blind pairs in comparison to some existing benchmarks. 
 \appendices

 \section{Proof of Proposition 1}  \label{app2}
 Let $(u,v)$ be a blind pair that is visible via $R^l_i$ using single reflection. Let $ d_{R^l_iu}$ and $ d_{R^l_iv}$ be the distances between $u$ and $R^l_i$, and $R^l_i$ and $v$, respectively. It is also given that $(u,v)$ is visible via $R^l_i$ and $R^m_j$ together using double reflections, where $R^l_i$ is a common subgroup that is used in both single and double reflections. Let $ d_{R^l_iR^m_j}$ and $ d_{R^m_jv}$ be the distances between $R^l_i$ and $R^m_j$, and $R^m_j$ and $v$, respectively. Additionally, we assume that $ d_{R^l_iR^m_j}>1$, $ d_{R^m_jv} > 1$ and the channel conditions for single and double-reflected communication are the same. Therefore, using triangular inequality, we can claim that
    \vspace{-2mm}
    \begin{equation}\label{lem}
        d_{R^l_iu} < d_{R^l_iR^m_j}+d_{R^m_jv}.
        \vspace{-1.5mm}
    \end{equation}
    Moreover, from \eqref{dr}, we have
    \vspace{-2mm}
\begin{align}
    & \!\!\!P|\mathbf{h}_{R^m_j} \!\!\times \!\mathbf{h}_{R^l_iR^m_j} \!\times \!\mathbf{h}_{uR^l_i}\!\times \!e^{j(\phi_{i,l}+\phi_{j,m})}|^2\!\!\times \! \rho_{\rm L}^3d_{R^m_jv}^{-\alpha}d_{uR^l_i}^{-\alpha}d_{R^l_iR^m_j}^{-\alpha}\nonumber \\
    & \leq  P|\mathbf{h}_{R^m_j} \times \mathbf{h}_{R^l_iR^m_j} \times \mathbf{h}_{uR^l_i}\times e^{j(\phi_{i,l}+\phi_{j,m})}|^2\nonumber \\ & \qquad \quad \times \rho_{\rm L}^2d_{R^m_jv}^{-\alpha}d_{uR^l_i}^{-\alpha}d_{R^l_iR^m_j}^{-\alpha} \;\; \left( \because \:\:0 < \rho <1 \right) \nonumber\\
    & \leq  P|\mathbf{h}_{R^m_j} \times \mathbf{h}_{R^l_iR^m_j} \times \mathbf{h}_{uR^l_i}\times e^{j(\phi_{i,l}+\phi_{j,m})}|^2\nonumber \\ & \;\;\; \times \rho_{\rm L}^2d_{uR^l_i}^{-\alpha}d_{R^l_iv}^{-\alpha} \; \text{(from \eqref{lem} and $ d_{R^l_iR^m_j} \!> \!1,d_{R^m_jv} \!> \!1$)}\label{ine}
    \vspace{-2mm}
\end{align}
Therefore, from \eqref{ine}, \eqref{dr} and \eqref{sr}, we can claim that
\begin{align}
    & \frac{P|\mathbf{h}_{R^m_j} \!\!\times \! \mathbf{h}_{R^l_iR^m_j} \!\! \times \! \mathbf{h}_{uR^l_i}\!\! \times \! e^{j(\phi_{i,l}+\phi_{j,m})}|^2\! \times \! \rho_{\rm L}^3d_{R^m_jv}^{-\alpha}d_{uR^l_i}^{-\alpha}d_{R^l_iR^m_j}^{-\alpha}}{\sigma^2} \nonumber\\
    & \leq \frac{P|\mathbf{h}_{R^m_j} \times \mathbf{h}_{R^l_iR^m_j} \times \mathbf{h}_{uR^l_i}\times e^{j(\phi_{i,l}+\phi_{j,m})}|^2\! \times \! \rho_{\rm L}^2d_{uR^l_i}^{-\alpha}d_{R^l_iv}^{-\alpha}}{\sigma^2}  \nonumber\\
    & \implies \gamma_{dr}\leq \gamma_{sr}
    \implies T^{l,m}_{i,j}(\gamma) \leq T^l_i(\gamma) \nonumber \\& \implies T^l_i(\gamma)-T^{l,m}_{i,j}(\gamma) \geq 0.\label{d_ratio}
\end{align}
Now, from \eqref{ef5} and \eqref{ef1}, we have 
\begin{align}
    & {\rm E^i_{ eff}}(l)- {\rm E^{i,j}_{eff}}(l,m)= \frac{T^l_i(\gamma)}{\frac{\beta \phi}{T^l_i(\gamma)} \times \Big(P+P_{\rm phase}(R^l_i)\Big)}\nonumber\\
    & \qquad \qquad-\frac{T^{l,m}_{i,j}(\gamma)}{\frac{\beta  \phi }{T^{l,m}_{i,j}(\gamma)} \times \Big(P+P_{\rm phase}(R^l_i)+P_{\rm phase}(R^m_j)\Big)}\nonumber\\
    & \!= \!\! \frac{1}{\beta \phi} \!\!\times \!\! \left \{\!\!\! \frac{\Big(P+P_{\rm phase}(R^l_i)+P_{\rm phase}(R^m_j)\Big)\Big(T^l_i(\gamma)\Big)^2}{\Big(\!P+P_{\rm phase}\!(R^l_i)+P_{\rm phase}(R^m_j)\Big)\Big(P+P_{\rm phase}(R^l_i)\Big)}\right. \nonumber \\ 
    &\;\;\;-\left. \frac{\Big(P+P_{\rm phase}(R^l_i)\Big)\Big(T^{l,m}_{i,j}(\gamma)\Big)^2}{\Big(P+P_{\rm phase}(R^l_i)+P_{\rm phase}(R^m_j)\Big)\Big(P+P_{\rm phase}(R^l_i)\Big)}\!\!\right\}\nonumber\\
    &\!=\!\! \frac{1}{\beta \phi} \!\!\times \!\!\left \{ \! \!\frac{\Big(P\!+\!P_{\rm phase}(R^l_i)\Big)\left(\Big(T^l_i(\gamma)\Big)^2\!-\!\Big(T^{l,m}_{i,j}(\gamma)\Big)^2\right)}{\!\Big(P\!+\!P_{\rm phase}(R^l_i)\!+\!P_{\rm phase}(R^m_j)\!\Big)\!\Big(P\!+\!P_{\rm phase}(R^l_i)\Big)}\right.\nonumber \\
    &\quad \left. +\frac{P_{\rm phase}(R^m_j)\Big(T^l_i(\gamma)\Big)^2}{\Big(P+P_{\rm phase}(R^l_i)+P_{\rm phase}(R^m_j)\Big)\Big(P+P_{\rm phase}(R^l_i)\Big)}\right\}\nonumber\\
    & \geq 0 \quad \left( \because \;\;\Big(T^l_i(\gamma)\Big)^2-\Big(T^{l,m}_{i,j}(\gamma)\Big)^2 \geq 0\;\; (\text{from}\; \eqref{d_ratio})\right. \nonumber \\& \qquad \qquad \quad \qquad \text{and} \;\; \left. P_{\rm phase}(R^m_j)\Big(T^l_i(\gamma)\Big)^2 \geq 0\right)\\
    & \implies  {\rm E^i_{ eff}}(l)- {\rm E^{i,j}_{eff}}(l,m) \geq 0.\label{fina}
\end{align}

    Therefore, we can conclude from \eqref{lem} that the transmitted signals from $u$ require a longer route for double reflections than for single reflection in order to reach $v$. As a result, due to having a longer path and substantial path loss, the achievable data rate at $v$ for double reflections are lower than the single reflection when utilizing a common subgroup. Hence, from \eqref{fina}, we can conclude that if a device pair is coverable by single and double reflections using a common subgroup, single reflection are more beneficial than double reflections which is quite intuitive.

\section{Proof of Proposition 2}  \label{app3}

Here we want to prove that double reflections may be more beneficial than single reflection in some specific scenarios. In this context, ${\rm E^s_{eff}}(t)$ for single reflection must be less than $ {\rm E^{i,j}_{eff}}(l,m)$ for double reflections. Therefore, from \eqref{ef5}  and \eqref{ef1}, we have
\vspace{-4mm}

\begin{align}
      {\rm E^{i,j}_{eff}}(l,m)\geq {\rm E^s_{eff}}(t) & \iff 
     \frac{T^{l,m}_{i,j}(\gamma)}{{\rm E^{i,j}_c}(l,m)} \geq\frac{T^i_s(\gamma)}{\rm {E^s_c}(i)} \nonumber \\ & \iff \frac{T^{l,m}_{i,j}(\gamma)}{T^i_s(\gamma)}\geq \frac{\rm {E^{i,j}_c}(l,m)}{\rm {E^s_c}(i)}.
\end{align}
Now, by using $ {\rm E^{i,j}_c}(l,m)$ and ${\rm E^s_c}(i)$ from \eqref{ec} and \eqref{ec1}, respectively, we obtain
\begin{align}
   \left(\frac{T^{l,m}_{i,j}(\gamma)}{T^i_s(\gamma)}\right)^2 \geq \frac{P+P_{\rm phase}(R^l_i)+P_{\rm phase}(R^m_j)}{P+P_{\rm phase}(R^t_s)}.\label{fi}
\end{align}

     From the stated condition $\rm (i)$, the required phase shift power for double reflections is always greater than the required phase shift power for single reflection. That is,
     \vspace{-3mm}
     \begin{align}\label{pow}
        & \!\!\! \Big(P+P_{\rm phase}(R^t_s)\Big) \leq \Big(P+P_{\rm phase}(R^l_i)+P_{\rm phase}(R^m_j)\Big) \\
        &\iff \frac{P+P_{\rm phase}(R^l_i)+P_{\rm phase}(R^m_j)}{P+P_{\rm phase}(R^t_s)} \geq 1 \label{pratio}.
        \vspace{-2mm}
    \end{align}
     
     Therefore, from \eqref{fi} and \eqref{pratio}, we can claim that
     \vspace{-2mm}
     \begin{align}
      & \left(\frac{T^{l,m}_{i,j}(\gamma)}{T^i_s(\gamma)}\right)^2 \geq 1 \nonumber \\
      &\iff \!\! T^{l,m}_{i,j}(\gamma)-T^t_s(\gamma) \!\geq 0  \left( \because T^{l,m}_{i,j}(\gamma)\!>0 \:\:\text{and}\:\: T^i_s(\gamma)>0\right) \nonumber\\
& \iff \dfrac{\splitfrac{P|\mathbf{h}_{R^m_j} \times \mathbf{h}_{R^l_iR^m_j} \times \mathbf{h}_{uR^l_i}\times e^{j(\phi_{i,l}+\phi_{j,m})}|^2}{\times \rho_{\rm L}^3d_{R^m_jv}^{-\alpha}d_{uR^l_i}^{-\alpha}d_{R^l_iR^m_j}^{-\alpha}}}{\sigma^2}  \nonumber\\
&\qquad \quad -\frac{P|\mathbf{h}_{R^t_sv}\times \mathbf{h}_{uR^t_i}\times e^{j\phi_{s,t}}|^2 d_{R^t_sv}^{-\alpha}d_{uR^t_s}^{-\alpha}\rho_{\rm L}^2}{\sigma^2} \geq 0.\label{ab1}
\vspace{-2mm}
\end{align}
      
Now, from the stated conditions $\rm (ii)$, we can show that
\vspace{-3mm}
\begin{align}
       &  \rho_{\rm L} d_{R^t_sv} d_{uR^t_s} \geq d_{R^m_jv} d_{uR^l_i} d_{R^l_iR^m_j}\nonumber \\ & \iff \rho^3_Ld_{R^m_jv}^{-\alpha}d_{uR^l_i}^{-\alpha}d_{R^l_iR^m_j}^{-\alpha} \geq \rho^2_Ld_{R^t_sv}^{-\alpha}d_{uR^t_s}^{-\alpha} .\label{dis11}
       \vspace{-3mm}
\end{align}

 Therefore, from \eqref{ab1} and \eqref{dis11}, we can claim that
 \vspace{-1mm}
 \begin{equation}\label{cin}
 |\mathbf{h}_{R^m_j} \!\times \! \mathbf{h}_{R^l_iR^m_j} \!\times \! \mathbf{h}_{uR^l_i}\times e^{j(\phi_{i,l}+\phi_{j,m})}|^2\!\geq \! |\mathbf{h}_{R^t_sv}\! \times \! \mathbf{h}_{uR^t_s}\! \times \! e^{j\phi_{s,t}}|^2.
\vspace{-1mm}
\end{equation}
That is, the power gain at the receiver end is greater in double reflections than in single reflection.

Moreover, from \eqref{cin}, we can conclude that if a device pair is visible via single and double reflection and it satisfies the stated conditions, then double reflection is more beneficial than single reflection.
 
\bibliographystyle{ieeetr}
\bibliography{ref}

\end{document}